\documentclass[prl,twocolumn]{revtex4}
\usepackage{amsmath,amssymb,lmodern}

\def \la {\langle}
\def \ra {\rangle}
\def \lla {\langle \langle}
\def \rra {\rangle \rangle}
\def \pa {\partial}

\def \eps {\epsilon}

\def \be  {\begin{equation}}
\def \ee  {\end{equation}}
\def \ba  {\begin{eqnarray}}
\def \ea  {\end{eqnarray}}
\def \baa {\begin{eqnarray*}}
\def \eaa {\end{eqnarray*}}
\def \bb  {\begin {thebibliography} }
\def \eb  {\end{thebibliography}}

\begin{document}
\title{A note on three-point functions of conserved currents}
\author{Alexander Zhiboedov}
\affiliation{Department of Physics, Princeton University, Princeton, NJ 08544}
\begin{abstract}
We find the form of three-point correlation functions of traceless symmetric conserved currents of arbitrary spin $\la j_{s_1} j_{s_2} j_{s_3} \ra$ in $d$-dimensional conformal field theory (CFT). These are fixed up to several constants by conformal symmetry and current conservation conditions. We present generating functionals for all structures in arbitrary $d$. In even dimensions we present an interpretation for each structure in terms of the corresponding free field. In odd dimensions $d>3$ an infinite number of structures is found which are not generated by known CFTs. 
\end{abstract}

\maketitle

\section{Introduction}

Three-point scattering amplitudes of massless particles in flat space are fixed up to a finite number of structures.
One can wonder whether the same is true for AdS three-point correlators, which are the AdS analogs of flat space scattering amplitudes.
The answer is known and given by the possible structures appearing in three-point functions of conserved currents in CFT living on the boundary. This is the well-known dictionary of AdS/CFT \cite{Mal97,GKP98,W} .

Recently there has been growing interest in the higher spin symmetric phase of the correspondence. Under scrutiny was Vasiliev theory  in $AdS_4$ (see e.g. \cite{Vasiliev:1999ba}) and it's dual large $N$ 3d free/critical $O(N)$ model \cite{Klebanov:2002ja, Sezgin:2002rt}. Three-point functions of conserved currents in the $AdS_4 / CFT_3$ were found to have a very simple form \cite{Giombi:2009wh, Giombi2011}. The natural question is whether similar formulas for three point functions exist in an arbitrary number of dimensions?

In this note we give an affirmative answer to this question and present explicit results for generating functionals of three-point functions of conserved currents. We use the notations of \cite{Costa2011mg} (with minor modifications). 

Recall that the idea is to introduce light-like polarization vector $\eps^{\mu}$, $\eps^2 =0$,
and then consider the object $j_{s} = j_{\mu_1 ... \mu_s} \eps^{\mu_1}...\eps^{\mu_s}$.

In the discussion below we do not consider the structures containing the epsilon tensor. These could only appear in $d=3,4$ \cite{Costa2011mg} and would be mentioned separately in the correspondent sections. Then the most general structure of correlation functions of conserved currents is given by 
\ba  \label{gengen} 
\la j_{s_1} j_{s_2} j_{s_3} \ra &=&\frac{\lla j_{s_1} j_{s_2} j_{s_3} \rra }{|x_{12}|^{d-2}|x_{13}|^{d-2}|x_{23}|^{d-2}}, \\
\lla j_{s_1} j_{s_2} j_{s_3} \rra &=& \sum_{i} c_{i} H_{12}^{h^{(12)}_i} H_{13}^{h^{(13)}_i} H_{23}^{h^{(23)}_i} V_1^{v^{(1)}_i} V_2^{v^{(2)}_i} V_3^{v^{(3)}_i} , \nonumber
\ea
where $V_{i}$ and $H_{i j}$ are conformal invariants (for their definition see appendix A) and
\ba \nonumber
h^{(j \ j+1)}+h^{(j \ j+2)}+ v^{(j)} = s_j.
\ea 
In all dimensions higher than three conformal invariants are independent. 


The problem is then to fix the coefficients $c_i$ that appear in front of all possible combinations by imposing current conservation conditions. The straightforward approach to the problem is very messy and complicated \cite{Osborn:1993cr}. 
One simplification is to work in the so-called embedding formalism. This was explained in  \cite{Costa2011mg}. However, even then imposing conservation of current is, in general, a rather cumbersome procedure.

The problem can be simplified drastically after noticing that for conserved currents $\pa_{\mu} j^{\mu}_{\ \mu_{2} ... \mu_{s}}$ is also a primary operator of the CFT. And the correlation function of $\la \pa.j_{s_1} j_{s_2} j_{s_3}\ra$ can
be again written as a known prefactor times the polynomial of the form discussed above but with $s_1 \to s_1 - 1$. Thus, taking the divergence of the current $j_{s} (x_i)$ can be thought as an operation in the space of polynomials of $V_i$'s and $H_{ij}$'s ${\cal D}_{i}: f(H_{i j}, V_i) \to g(H_{i j}, V_i)$.
This point of view was taken in \cite{Maldacena:2011jn}. 

Moreover, since the explicit form of the divergence operator is known one can easily check that ${\cal D}_{i}$ can be at most a third order differential operator and moreover, it cannot contain the $\frac{\pa}{\pa H_{i+1, i+2}}$ derivatives.
By writing the most general form of the operator and mapping it to the result of taking the divergence of the most general structure from (\ref{gengen}) one can fix it completely. The result is given in the appendix A.

\section{Solution}

Given the explicit form of ${\cal D}_{i}$ one can try to find the most general solution of the equation
\be\label{diveq}
{\cal D}_{i} \lla j_{s_1} j_{s_2} j_{s_3} \rra (H_{i j}, V_i)  = 0.
\ee
For $d=3$ the solutions corresponding to a free boson and free fermion were found in 
\cite{Giombi:2009wh}. In \cite{Maldacena:2011jn} using the light cone limit it was proved that these are
the only solutions. 

To proceed to higher dimensions first notice that the solutions of (\ref{diveq}) do exist in an arbitrary $d$ and for arbitrary $s_i$'s. This follows from the existence of free boson and fermion theories in arbitrary dimensions. Moreover, the solution in $d=3$ is known to have a factorized form
\be
\lla j_{s_1} j_{s_2} j_{s_3} \rra_{d=3} = \prod_{i =1}^{3} \beta( V_i )  \prod_{i<j ; 1}^3 \gamma (H_{i j}) 
\ee
This serves as the natural starting point to guess the answer for an arbitrary $d$. One can check that, indeed, the solution of this form exists. 

In $d=3$ the conformal invariants are not independent, namely there is a relation between them
\be\label{relation}
-\frac{1}{2}H_{12} H_{13} H_{23} = \left( V_1 V_2 V_3 + \frac{1}{2} \left[ V_1 H_{23} + V_2 H_{13} + V_3 H_{12} \right]  \right)^2 .
\ee

It is also instructive to recall the behavior of free field correlators with all minus components
$j_{-...-}$ when $x_{i j}^{+} \to 0$ \footnote{We use the metric $d s^2 = d x^+ d x^- + \sum_{i=1}^{d-2} d x_{i}^2$.}. We expect conserved currents to have a schematic form $j_{- s} = \psi_{-j}  \overleftrightarrow{\pa}_{-}^{s- 2j} \psi_{-j}$ where $\psi_{-j}$ is a field of spin $j$ and the propagator is by Lorentz invariance 
\be
\la\psi_{-j} (x) \psi_{-j}(0) \ra \propto (x^+)^{2j} .
\ee

This guarantees that correlation functions of currents built using field $\psi_{-j}$ will also have the limit
\be \label{limitb}
\lim_{x^+_{ij} \to 0 }\lla j_{-s_1} j_{-s_2} j_{-s_3} \rra \propto (x^+_{ij})^{2j} .
\ee

To capture this limit it is useful to introduce the following combinations of conformal invariants
\ba \label{deflam}
\Lambda_1 &=& V_1 V_2 V_3 + {1 \over 2} \left[ V_1 H_{23} + V_2 H_{13} + V_3 H_{12} \right], \\ \nonumber
\Lambda_2 &=& H_{12} H_{13} H_{23} .
\ea

If we choose $\eps_{i}^{\mu} = \eps^{-}$ then the relation (\ref{relation}) holds 
in any number of dimensions $\Lambda_2 |_{\eps_{i}^{\mu} = \eps^{-}} = - 2 \Lambda_1^2  |_{\eps_{i}^{\mu} = \eps^{-}}$. 
Moreover, as one can check in this case 
\ba
\Lambda_1  |_{\eps_{i}^{\mu} = \eps^{-}} &=&  {1 \over 4} {x_{12}^{+} x_{13}^{+} x_{23}^{+} \over  x_{12}^2 x_{13}^2 x_{23}^2 } (\eps^{-})^3.
\ea

In this manner the most general prefactor that captures the limit (\ref{relation}) is $\sum_{i=0}^{[j]} c_{i} \Lambda_1^{2 j - i} \Lambda_2^{i}$.

Below it will be useful to introduce the following function
\be
{\cal G} (\alpha) =  \prod_{i =1}^{3} e^{V_i}  \prod_{i<j ; 1}^3 {}_0 F_{1}(\alpha, - {H_{i j} \over 2})
\ee
where $ {}_0 F_{1}$ is a usual $(0,1)$ hypergeometric function. 


The simplest solution that has factorized form and limit behavior (\ref{limitb}) with $j=0$ takes the form
\be \label{scalar}
\lla j_{s_1} j_{s_2} j_{s_3} \rra_{d}^{j=0} = {\cal G} ({d \over 2} - 1).
\ee

Let us explain how one should understand this formula. To extract the three-point function for particular $s_i$ one needs to pick up the $\eps_1^{s_1} \eps_2^{s_2} \eps_3^{s_3}$ component in the expansion of the RHS of (\ref{scalar}). Formulas below should be understood in the same manner.

The answer for $j={1 \over 2}$ should contain a $\Lambda_1$ factor. Again trying the factorized ansatz for the generating functional one finds the solution
\be
\lla j_{s_1} j_{s_2} j_{s_3} \rra_{d}^{j={1 \over 2}} = {\cal G} ({d \over 2}) \Lambda_1 .
\ee

The first case where both $\Lambda_1$ and $\Lambda_2$ appear is the case of $j=1$. Again we find that the solution exists and is given by
\be
\lla j_{s_1} j_{s_2} j_{s_3} \rra_{d}^{j=1} = {\cal G} ({d \over 2} + 1) \left[ \Lambda_1^2 +{\Lambda_2  \over 2  (d-2)}  \right].
\ee
As a trivial check notice that it becomes zero when $d=3$ due to (\ref{relation}). 

After considering simplest cases we can present the general integer or half-integer spin $j$ solution. The spin here is defined by the light cone limit (\ref{limitb}) and not by the existence of some actual elementary field in an abstract Lagrangian. 

As one can easily check using the operator from appendix A the spin $j$ solution takes the form

\ba \label{general} \nonumber
&&\lla j_{s_1} j_{s_2} j_{s_3} \rra_{d}^{j} ={\cal G} ({d \over 2}+2 j - 1) \Lambda_1^{2 j} \\
 &&\times {}_2 F_{1} ({1 \over 2} - j , - j, 3 - {d \over 2} - 2 j, -  {1 \over 2} {\Lambda_2 \over \Lambda_1^2} ).
\ea

Let us comment on some general features of the solution (\ref{general}). First, in the light cone limit it has behavior (\ref{limitb}).
Secondly, each spin $j$ structure contributes only to the correlation functions with $s_{i} \geq 2 j$ \footnote{ This is in accord with the Weinberg-Witten rule \cite{Weinberg:1980kq}: the existence of the field of spin $j$ charged under symmetry generated by gauge invariant operator $j_s$ implies $s \geq 2 j$.}.
Thirdly, for $d=3$ all structures with $j> {1 \over 2}$ are zero.

Thus, we get that the most general form of the three-point functions in general $d$ is given by
\be \label{result}
\lla j_{s_1} j_{s_2} j_{s_3} \rra_{d} =\sum_{ j = 0}^{{\rm min}[s_1, s_2, s_3]/2} c_j \lla j_{s_1} j_{s_2} j_{s_3} \rra_{d}^{j} 
\ee
where $j$ takes integer and half-integer values. Notice that we have in total ${\rm min}[s_1, s_2, s_3] + 1$ structures for $d>3$. In \cite{Costa2011mg} this number was conjectured as the number of possible structures. The uniqueness of the solutions presented above for $j=0$ and $j ={1 \over 2}$  is proved in the appendix B. The fact that (\ref{result}) is a complete set of solutions is also supported by the calculation of particular examples. 

For completeness an appendix C we present the result for three-point functions of stress tensors.

\section{Interpretation}


Above we introduced the basis of all three point functions governed by the limit (\ref{limitb}) and inspired by the structure of free field theories.
One can wonder what is the precise relation between the solutions above and the three point functions that appear in free field theories.

Given a free field of spin $j'$ the three point functions of conserved currents in this theory by construction above will generate the $j'$-structure.
However, a priori the structures with $j>j'$ could be also generated.

It is easy to check that for three point function of stress tensors each $j$ solution corresponds to free field: $j=0$ being a free scalar, $j = {1 \over 2}$ a free fermion and $j=1$ in even dimensions $d>4$ is generated by the free theory of ${d \over 2} - 1$ form (see \cite{Anselmi:1999bb, Buchel:2009sk}).

One can wonder whether this holds true for general spins, namely given a free field theory it generates three point functions with given $j$ considered above. We do not have a proof of this statement but below we interpret different solutions {\it assuming} that this is true, namely {\it each unitary free field theory generates only one $j$ structure}. It would be interesting to check it.

\subsection{$d=3$}

For $d=3$ the two structures correspond to the theory of a free boson ($j=0$)
\be
\lla j_{s_1} j_{s_2} j_{s_3} \rra_{3}^{{\rm boson}} = \prod_{i =1}^{3} e^{V_i}  \prod_{i<k ; 1}^3  \cos (\sqrt{ 2 H_{i k} })
\ee
 and a free fermion ($j={1\over 2}$)
\be
\lla j_{s_1} j_{s_2} j_{s_3} \rra_{3}^{{\rm fermion}} = \prod_{i =1}^{3} e^{V_i}  \prod_{i<k ; 1}^3  \sin (\sqrt{ 2 H_{i k} }) 
\ee
The corresponding formulas were found in \cite{Giombi:2009wh, Giombi2011}. 

For spins satisfying the triangle rule $s_{i} \leq s_{i+1} + s_{i+2}$ there is an additional structure containing the epsilon tensor discussed in \cite{Maldacena:2011nz, Giombi2011}. The integral expression for it could be found in the appendix B of \cite{Maldacena:2011jn}.

\subsection{$d=4$}

For $d=4$ the answer can be written as follows \footnote{For any even $d$ we can use the relation $$ {}_0 F_{1}({d \over 2}+2 j - 1, -{x \over 2 }) \propto {J_{{d -4\over 2}+2 j } (\sqrt{2 x}) \over x^{j + {d -4\over 4} } }.$$ }
\ba \label{dfour} \nonumber
\lla j_{s_1} j_{s_2} j_{s_3} \rra_{4}^{j} &=& \prod_{i =1}^{3} e^{V_i}  \prod_{i<k ; 1}^3 {J_{2 j} (\sqrt{ 2 H_{i k} }) \over H_{i k}^{j}} \Lambda_1^{2 j} \\
&&  {}_2 F_{1} ({1 \over 2} - j , - j, 1 - 2 j, -  {1 \over 2} {\Lambda_2 \over \Lambda_1^2} )
\ea
where $J_{2 j}$ is an ordinary Bessel function. It is known that there is an infinite set of free CFT's in $d=4$ given by $(j,0)$ self-dual tensors  \cite{Siegel:1988gd, Minwalla:1997ka}. Thus, we conjecture that  $j$ labels the rank of the self-dual tensor. The equation of motions and conserved current operators for these fields were neatly described in \cite{Gelfond:2006be}. From there it is clear that the correlation functions of currents in these theories will have the light cone limit as we described as well as the fact that conserved currents for rank $j$ field exist starting from spin $2 j$.

The equivalent form of (\ref{dfour}) as well as the identification with correspondent free fields were also presented in \cite{Stanev:2012nq}.

For completeness we present here a formula for the possible three point functions involving the epsilon tensor that were found in \cite{Stanev:2012nq}. A generating functional equivalent to the one presented in \cite{Stanev:2012nq} is
\ba \label{dfourodd} \nonumber
\lla j_{s_1} j_{s_2} j_{s_3} \rra_{4, {\rm odd } }^{j} &=&  S \prod_{i =1}^{3} e^{V_i}  \prod_{i<k ; 1}^3 {J_{2 j} (\sqrt{ 2 H_{i k} }) \over H_{i k}^{j}} \Lambda_1^{2 j} \\
&&  {}_2 F_{1} ({1 \over 2} - j , - j, - 2 j, -  {1 \over 2} {\Lambda_2 \over \Lambda_1^2} )
\ea
where $S$ is the unique conformal invariant involving the epsilon tensor which is defined in appendix A. According to \cite{Stanev:2012nq} $j$ goes from $0$ to ${ \min [s_1, s_2, s_3] - 1\over 2}$ taking integer and half-integer values. In this way $\min [s_1,s_2,s_3]$ number of structures is generated.

\subsection{Even $d>4$}

In this case we have again an infinite number of free fields labeled by their representation \cite{Siegel:1988gd, Minwalla:1997ka}. We expect that our label $j$ corresponds to the representation label for free field theories in even dimensions analogously to the case $d=4$.

\subsection{Odd $d>3$}

The situation in odd dimensions is more interesting. There are no free fields except a free scalar ($j=0$) and free fermion ($j = {1 \over 2}$). Nevertheless, we have an infinite number of different structures. The natural question is: where do they come from \footnote{The question about the origin of the third structure in the three-point function of stress tensors was pointed to us by Rob Myers.}?

We would like to point out the class of free CFT's that could be relevant for this question. 
The scenario we have in mind was considered in the \cite{ElShowk:2011gz} in the context of $d$-dimensional Maxwell theory. In their case a non-unitary CFT contained operator $F_{\mu \nu}$ which satisfies
\be
\Box F_{\mu \nu} = 0,
\ee
however, compared to the usual free field it is {\it not a primary operator}.

This theory being free has higher spin symmetry and, thus, higher spin currents. This is easy to see since the transformation $[Q_{s}, F_{\mu \nu}] = \pa^{s-1} F_{\mu \nu}$ is a symmetry of correlation functions \footnote{As well as $[Q_{s}, A_{\mu}] = \pa^{s-1} A_{\mu}$ by the same token.}. 


One can wonder what are the correlation function of conserved currents in this exotic theory and which structures do they generate. We do not explore
this question in the present note.

More general scenario would be to consider a field $\psi$ such that
\ba
\Box \psi = 0 \\
K_{\mu} \psi (0) \neq 0
\ea
where $K_{\mu}$ is the special conformal generator. 

One can also wonder whether there exist unitary conformal theories  (non-Lagrangian) in odd dimensions such that they have correlation functions of conserved currents being given by $j > {1 \over 2}$ structure. We do not know the answer to this question. The bootstrap approach of \cite{Maldacena:2011jn} could be useful to clarify this point.

\section{Case of one conserved current}

More generally, one can consider the problem of finding three-point function $\la j_{s_1} j_{\tau_2, s_2}  j_{\tau_3, s_2} \ra$ where only $j_{s_1}$ is conserved ($\tau = \Delta - s$ is the twist of the operator). To analyze this case we present at the end of the appendix A an extension of the differential operator that can be used for this general case.  

From the form of the differential operator it is clear that in the case when $\tau_2 = \tau_3 = \tau$ we can use the solutions found for the case of conserved currents, namely we can write the following set of solutions
\ba \nonumber
\la j_{s_1} j_{\tau, s_2}  j_{\tau, s_3} \ra &=&\frac{\lla j_{s_1} j_{\tau, s_2}  j_{\tau, s_3} \rra }{ |x_{12}|^{d-2}|x_{13}|^{d-2}|x_{23}|^{2 \tau - (d-2)}}, \\ \nonumber
\lla j_{s_1} j_{\tau, s_2}  j_{\tau, s_3} \rra &=& \sum_{l = 0}^{{\rm min}[s_2 , s_3]}  \sum_{k = 0}^{{\rm min}[s_1, s_2 -l , s_3 - l]}  c_{k l} \\ 
&&H_{23}^l \lla j_{s_1} j_{s_2 - l} j_{s_3 - l} \rra^{{k \over 2} } ,
\ea
where each summand is annihilated by ${\cal D}_{1}$. Notice that in this case a total number of possible structures is equal to
\be
\frac{1}{2}({\rm min}[s_1,s_2,s_3] + 1 ) (2 \ {\rm min}[s_2,s_3] - {\rm min}[s_1,s_2,s_3] + 2)
\ee 
and grows quadratically with spin. We do not prove that these are {\it all} possible structures.
However, for all particular examples we found that these are all possible structures \footnote{For example, we checked that there are $42$ solutions for $s_1 = 6$, $s_2 = 8$, $s_3=10$.}. 

\section{Three point function of stress tensors from the gravity dual}

It is also instructive to present the result for the three point function of stress tensors
computed using gravitational action in AdS
\be
S = {1 \over 2 \kappa_{d+1}^2 } \int d^{d+1} x \sqrt{g} \left( R - 2 \Lambda \right).
\ee

 This problem was solved in \cite{Arutyunov:1999nw}. Here we present the result of  \cite{Arutyunov:1999nw} using the solutions introduced above.
It makes the answer more transparent and easier to understand. The relevant three point functions are written in appendix C. 
We get 
\ba
\lla T T T\rra^{gravity} &=& a_{j=0} \lla j_{2} j_{2} j_{2} \rra_{d}^{j=0} \\ \nonumber
&+& a_{j = {1 \over 2}}\lla j_{2} j_{2} j_{2} \rra_{d}^{j={1 \over 2}} + a_{j=1}\lla j_{2} j_{2} j_{2} \rra_{d}^{j=1}
\ea
where
\ba
a_{j=0} &=&  {d -2 \over 2 (d-1)^2} s_{d} \\
a_{j = {1 \over 2}} &=& s_{d}  \\ \label{jonesol}
a_{j=1} &=& d \ s_{d}
\ea
and $s_{d} = - {128 d^2 \Gamma(d) \over (d-1) \pi^d \kappa_{d+1}^2}$. The formulas above are valid for $d \geq 4$.

For $d=4$ the answer is very well known and corresponds to the computation in strongly coupled large $N$ ${\cal N}=4$ SYM.

For $d=6$ it corresponds to the three point function in large $N$ $A_{N-1}$ $(2,0)$ SCFT \cite{Seiberg:1997ax, Aharony:1999ti}.

For $d=5$ interacting fixed points involving supergravity dual were described in \cite{Seiberg:1996bd, Brandhuber:1999np, Bergman:2012kr}. It is interesting to notice that the structure $j=1$ is generated in this case. It indicates non weakly coupled nature of these theories and serve as an example of the appearance of the $j=1$ structure in unitary interacting $5d$ SCFTs.

\section{Conclusions} 

The main result of this note are the formulas (\ref{general},\ref{result}). Very likely they comprise
all possible structures that can appear in the three-point functions of conserved currents. This is supported
both by the general counting arguments of \cite{Costa2011mg} as well as by ``experimental'' data of particular examples.
In even dimensions each structure originates from a free field. In odd dimensions the situation is 
less clear. Namely we found an infinite number of structures which do not originate from any known 
CFT.

 We see the following possible applications of these results.

First, they could be useful in extending the argument of \cite{Maldacena:2011jn} to higher dimensions. 
Secondly, this analysis could be thought of as the computation of possible on-shell structures that could appear in the higher spin gauge theories in AdS \cite{Vasilev:2011xf,Joung:2011ww}. 
Thirdly, they could be useful in looking for weakly coupled interacting fixed points in higher dimensions in the spirit of \cite{Maldacena:2012sf,Giombi:2011kc}. 
Fourthly, this is an exercise in computing tree-level correlators of operators in any weakly coupled CFT.

It would be interesting to clarify the origin of the $j> {1 \over 2}$ structures in odd dimensions $d>3$ as well as to understand the structure of correlation functions in the theories of the type considered in \cite{ElShowk:2011gz}.


Another obvious extension is to consider the three-point functions of the currents in more general representations of the Lorentz group.

{\bf Note:} We became aware of \cite{Stanev:2012nq} while this paper was in preparation. The result for three-point functions in $d=4$ without the epsilon tensor found in \cite{Stanev:2012nq} agrees with the formula (\ref{dfour}) obtained in the present note. 

To summarize, the results of \cite{Stanev:2012nq}, results of this note as well as results from previous papers comprise a complete classification of all possible structures for three point functions of symmetric traceless conserved currents in arbitrary $d$.

{\it Acknowledgments:} We are grateful to B. Basso, D. McGady for useful discussions. We thank A. Dymarsky, D. McGady and B.Safdi for comments on the manuscript.  We thank D. Poland for providing us with the code to test the ideas considered in the note. We are especially grateful to J.Maldacena for discussions on this and related topics, comments, suggestions and reading of the manuscript.

The work of AZ was supported in part by the US National Science Foundation under Grant No. PHY-0756966 and by the Department of Energy under Grant No.\#DEFG02-91ER40671.

\section{Appendix A}

The conformal invariants used in the main text are given by the following formulas
\ba
V_{i} &=& \frac{ (\eps_i x_{i , i+2})}{x_{i,i+2}^2} - \frac{(\eps_i x_{i, i+1})}{x_{i,i+1}^2}, \\
H_{i j} &=& {(\eps_i \eps_j) x_{ij}^2 - 2 (\eps_i x_{ij})  (\eps_j x_{ij}) \over  x_{ij}^4}, \\ \nonumber
x_{i j} &=& x_{i} - x_{j}.
\ea
where under conformal mapping $x \to \hat{x}$ polarization tensors transform as $\eps^{\mu}_{i} \to \hat{\eps}^{\mu}_{i} = {\pa \hat{x}^{\mu} \over \pa x^{\nu}}|_{x = x_i} \eps^{\nu}_{i}$.

In four dimensions there is an additional unique conformal invariant that contains the epsilon tensor $\eps_{\mu \nu \rho \sigma}$. It can be written as follows
\ba \nonumber
S &=& { \eps_{\mu \nu \rho \sigma} \over x_{12}^2 x_{13}^2 x_{23}^2 } \left[ \eps_2^{\mu} \eps_3^{\nu} x_{13}^{\rho} x_{23}^{\sigma}  (\eps_1 x_{1 3}) - \eps_1^{\mu} \eps_3^{\nu} x_{13}^{\rho} x_{23}^{\sigma}  (\eps_2 x_{2 3}) \right. \\ \nonumber
&+& \left. {1 \over 2} \eps_1^{\mu} \eps_2^{\nu} \eps_{3}^{\rho} x_{23}^{\sigma} x_{13}^2 - {1 \over 2} \eps_1^{\mu} \eps_2^{\nu} \eps_{3}^{\rho} x_{13}^{\sigma} x_{23}^2\right].
\ea

Here we present the form of the divergence differential operator. We write only ${\cal D}_1$, other operators
can be obtained by a cyclic change. Since the operator is rather cumbersome we split it into
three parts according to the number of derivatives contained in it
\ba \label{firstpd}
&&{\cal D}_1 = {{\cal D}_1^{(3)} \over d-2 } + {\cal D}_1^{(2)} + (d-2) {\cal D}_1^{(1)} \\ \nonumber
&&{\cal D}_1^{(1)} = V_2 \pa_{H_{12}} - V_3 \pa_{H_{13}} \\ \nonumber
&&{\cal D}_1^{(2)} = 2 H_{12} V_2 \pa_{H_{12}}^2 +  4 H_{13} V_2 \pa_{H_{13}} \pa_{H_{12}}\\ \nonumber
&& - (H_{23} + 2 V_2 V_3 ) \pa_{V_{2}} \pa_{H_{13}} + V_2 \pa_{V_{1}} \pa_{V_{2}}\\ \nonumber
&&+ (H_{12} + 2 V_1 V_2 ) \pa_{V_{1}} \pa_{H_{12}} - [2 \to 3]  \\ \nonumber
&&{\cal D}_1^{(3)} = 8 H_{12} (H_{13} V_2 - \Lambda_1) \pa_{H_{12}}^2 \pa_{H_{13}} \\ \nonumber
&&+ 4 \Lambda_1  \pa_{V_1} \pa_{V_3} \pa_{H_{12}}+ 4 H_{13} V_{2}  \pa_{V_1} \pa_{V_2} \pa_{H_{13}} \\ \nonumber
&&+ 2 H_{12} ( H_{23}+ 2 V_2 V_3)  \pa_{V_3} \pa_{H_{12}}^2 \\ \nonumber
&& +2 H_{12} (H_{12} + 2 V_1 V_2) \pa_{V_1} \pa_{H_{12}}^2 \\ \nonumber
&& +(H_{12} + 2 V_1 V_2) \pa_{V_1}^2 \pa_{V_2}  - [2 \to 3],
\ea
where $\Lambda$'s were defined in (\ref{deflam}) and $ [2 \to 3]$ stands for the anti-symmetrization of the whole expression with respect to $2$ and $3$.
One can check that (\ref{general}) is annihilated by this differential operator.

In the case when we want to impose conservation of a current inside the three-point function $\la j_{s_1} j_{\tau_2, s_2}  j_{\tau_3, s_2} \ra$ where
$\tau = \Delta - s$ is the twist of the field we get an additional piece in the differential operator $$(\tau_2 - \tau_3) \left[ \delta {\cal D}^{(1)}+{\delta {\cal D}^{(2)} \over d-2} \right]$$

\ba \label{secondpd}
&&\delta {\cal D}_1^{(1)} = \pa_{V_{1}} - V_2 \pa_{H_{12}} - V_3 \pa_{H_{13}} \\ \nonumber
&&\delta {\cal D}_1^{(2)} = V_1 \pa_{V_1}^2 - 4 \Lambda_1 \pa_{H_{12}} \pa_{H_{13}} \\ \nonumber
&&+ 2 \pa_{V_1} \left(  H_{12}  \pa_{H_{12}} + H_{13}  \pa_{H_{13}}\right)  \\ \nonumber
&&- 2 \left(V_2 H_{12} \pa_{H_{12}}^2 + V_3 H_{13} \pa_{H_{13}}^2 \right)
\ea

The total differential operator is the sum of (\ref{firstpd}) and (\ref{secondpd}).

These expressions for the differential operator are not particularly elegant. One can wonder if these can be simplified with the different choice of the conformal basis for the three-point function.

\section{Appendix B}

Here we would like to argue that the solutions that we obtained above
are unique for $j=0$ and $j = {1 \over 2}$. We are following the same route as in \cite{Maldacena:2011jn}.

Recall that the guiding idea for finding the solutions was their behavior in the particular limit $\lim_{x_{i j}^+ \to 0} j_{-...-}(x_i)  j_{-...-}(x_j)$. 

In this limit correlation functions simplify and one can hope to fix them in a easier way. 

We would like to impose the conservation of current at point $x_1$ while taking the limit described
above for point $x_{23}^+$. In terms of $V$'s and $H$'s this limit corresponds to imposing the conservation condition in the space of polynomials around the point
\ba \label{subspace} \nonumber
H_{23} &=& 0, \\
V_1 &=& - {1 \over 2} \left[ {H_{13} \over V_3} + {H_{12} \over V_2} \right].
\ea

Then we arrange different structures according to their behavior around this point. The analysis is especially simple in the cases of $j=0$ and $j = {1 \over 2}$. 

\subsection{$j = 0$}

In this case we have
\ba \nonumber
\lla j_{s_1} j_{s_2} j_{s_3} \rra &=&  \sum_{a = 0}^{s_1} c(a) H_{12}^a H_{13}^{s_1
 -a} \\ \nonumber 
&&V_2^{s_2 - a } V_3^{s_3 - s_1 + a} 
\ea
where powers were fixed to give the correct spin. Now we act on this function with ${\cal D}_1$ to get
\be
{c(a+1) \over c(a) } = {(s_1 - a) (s_1 +{d - 4 \over 2} - a) (s_2 + a + {d - 2 \over 2}) \over (a+1) (a+{d - 2 \over 2}) (s_1 + s_3  +{d - 4 \over 2} -a )}.
\ee

\subsection{$j = {1 \over 2}$}

In this case we have
\ba \nonumber
\lla j_{s_1} j_{s_2} j_{s_3} \rra &=& \Lambda_1 \sum_{a = 0}^{s_1 - 1} c(a) H_{12}^a H_{13}^{s_1 - 1 -a} \\ \nonumber 
&&V_2^{s_2 - 1 - a } V_3^{s_3 - s_1 + a} 
\ea
where powers were fixed to give the correct spin. Again we act on this function with ${\cal D}_1$ to get
\be
{c(a+1) \over c(a) } = {(s_1 - 1 - a) (s_1 +{d - 4 \over 2} - a) (s_2 + a + {d - 2 \over 2}) \over (a+1) (a+{d - 2 \over 2} + 1) (s_1 + s_3  +{d - 4 \over 2} - 1 -a )}.
\ee

\subsection{$j = 1$}

Starting from $j=1$ the limit is not enough to fix the structure completely for $d>3$. It is related to the fact that we can have several $\Lambda$ terms ($\Lambda_1^2$ and $\Lambda_2$ for $j=1$) with different series in front of each term. We will get the recursion relation again, however, it is not enough to fix the solution completely. 

However, for the ansatz 
\ba \nonumber
\lla j_{s_1} j_{s_2} j_{s_3} \rra &=&(\Lambda_1^2 + \gamma \Lambda_2) \sum_{a = 0}^{s_1 - 2} c(a) H_{12}^a H_{13}^{s_1 - 2 -a} \\ \nonumber 
&&V_2^{s_2 - 2- a } V_3^{s_3 - s_1 + a} 
\ea
it is possible to check that the solution is unique and is given by the one in the text. 


For $j>1$ we will get even more $\Lambda$ terms and additional analysis would also be needed to argue the uniqueness
of the solution.

\section{Appendix C}

Here we present the most general structure for the three-point function of stress tensors as it comes from the $j=0,{1 \over 2},1$ structures
\ba \nonumber
\lla j_{2} j_{2} j_{2} \rra_{d}^{j=0} &=& \Lambda_1 \left[  \Lambda_1  - (d+2) V_1 V_2 V_3 \right] - {d \over (d-2)^2} \Lambda_2 \\ \nonumber
&+& {1\over 8} (d+2)(d+4) V_1^2 V_2^2 V_3^2 \\ \nonumber
&+&  {d+2 \over 2 (d-2)}  \left(V_1 V_2 H_{13} H_{23} +{\rm cyclic} \right),\\ \nonumber
\lla j_{2} j_{2} j_{2} \rra_{d}^{j = {1 \over 2}} &=& \Lambda_1 \left[  \Lambda_1  -{(d+2) \over 2} V_1 V_2 V_3 \right],\\ \nonumber
\lla j_{2} j_{2} j_{2} \rra_{d}^{j = 1} &=& \Lambda_1^2 +{\Lambda_2  \over 2  (d-2)} .
\ea
These are the only structures that contribute to the three-point function of stress tensors for $d \geq 4$. 

In $d=3$ there is an additional structure that contains the epsilon tensor. It was originally found in momentum space in \cite{Maldacena:2011nz} (see \cite{Giombi2011} for it's position space form). In $d=4$ structures containing the epsilon tensor are not allowed by permutation symmetry.

For completeness we also present a relation between the structures above and answers obtained from free theories.

For a real scalar and a Dirac fermion in arbitrary $d$ we get \cite{Erdmenger:1996yc}
\ba \nonumber
\lla j_{2} j_{2} j_{2} \rra_{d}^{scalar} &=& - {8 d^2 (d-2)^2 \Gamma({d \over 2})^3 \over (d-1)^3 \pi^{{3d \over 2}}} \lla j_{2} j_{2} j_{2} \rra_{d}^{j=0} \\ \nonumber
\lla j_{2} j_{2} j_{2} \rra_{d}^{fermion} &=& - {8 d^2 2^{[{d \over 2}]} \Gamma({d \over 2})^3 \over \pi^{{3d \over 2}}}\lla j_{2} j_{2} j_{2} \rra_{d}^{j = {1 \over 2}} ,
\ea
in even $d$ we can also have a theory of $({d - 2\over 2})$-form \cite{Buchel:2009sk}
\ba \nonumber
\lla j_{2} j_{2} j_{2} \rra_{d}^{({d - 2\over 2})-form} =- {16 d^3 (d-2) \Gamma({d \over 2}) \Gamma(d-1) \over (d-3) \pi^{{3d \over 2}}} \\ \nonumber
 \lla j_{2} j_{2} j_{2} \rra_{d}^{j = 1}.
\ea

\end{document}